\documentclass[pra,twocolumn]{revtex4}

\usepackage{graphicx}

\newcommand{\ket}[1]{\mbox{\ensuremath{|#1\rangle}}}

\begin{document}

\title{Eliminating Spectral Distinguishability in Ultrafast Spontaneous Parametric
Down-conversion}

\author{Hou Shun Poh$^1$}
\author{Jiaqing Lim$^1$}
\author{Ivan Marcikic$^2$}
\author{Ant\'{\i}a Lamas-Linares$^1$}
\author{Christian Kurtsiefer$^1$}

\affiliation{$^1$Centre for Quantum Technologies and Department of Physics, National University of Singapore, 3 Science Drive 2, Singapore 117543\\
$^2$Temasek Laboratories, National University of Singapore, Singapore, 117508}

\date{\today}

\begin{abstract}
Generation of polarization-entangled photon pairs with a precise timing
through down-conversion of femtosecond pulses is often faced with a
degraded polarization entanglement quality. In a previous experiment we have
shown that this degradation is induced by
spectral distinguishability between the two decay paths, in accordance with
theoretical predictions. Here, we present an experimental study of
the spectral compensation scheme proposed and first implemented by Kim
\textit{et al.}~\cite{kim:02}. By measuring the joint spectral properties of the
polarization correlations of the photon pairs, we show that the spectral
distinguishability between the down-converted components is eliminated. This
scheme results in a visibility of 97.9\,$\pm$\,0.5\% in the complementary
polarization basis without any spectral filtering.

\end{abstract}


\maketitle

Spontaneous parametric down-conversion (SPDC) has been widely used to generate
entangled photons required in various quantum information
protocols~\cite{bouwmeester:01}. In some experiments, light from
continuous-wave lasers is used to pump the SPDC process~\cite{kwiat:95,
  burnham:70, klyshko:89}. These sources can be very bright and provide photon
pairs in maximally entangled states with high fidelity in various degrees of
freedom \cite{brendel:92, kwiat:99}, making them suitable for applications
such as quantum key distribution~\cite{jennewein:00} and fundamental tests of
quantum physics (e.g. tests of Leggett models~\cite{groblacher:07, branciard:08}).

For experiments which require photon pairs to exhibit tight localization in
time~\cite{bouwmeester:97, linares:02}, or for preparation of entangled states
between more than two photons~\cite{gaertner:03, goebel:03}, the SPDC process
needs to be pumped by ultrafast optical pulses. Such sources often exhibit a
reduction in the quality of polarization entanglement arising from spectral
distinguishability of the possible decay paths. This has been addressed
theoretically~\cite{keller:97, grice:97, grice:98}; more recently,
experiments investigating the underlying phenomenom have been
performed~\cite{atature:99, kim:05, poh:07, wasilewski:06,
  avenhaus:09}. Various techniques are implemented to eliminate spectral
distinguishability: they range from specific tailoring of the
down-conversion medium~\cite{erdmann:00}, double-pass configuration of the
pump beam~\cite{hodelin:06} to interferometric setups~\cite{branning:99}.

In previous work, we have shown that the wider spectral
distribution of ordinarily (\textit{o}) polarized down-converted light in
comparison with the extraordinary (\textit{e}) component translates into a
spectral distinguishability between the two decay paths for type-II
SPDC~\cite{poh:07}.
When only the polarization degree of freedom is considered, this spectral
distinguishability reduces the purity of a state and thus the
entanglement quality. Typically, strong spectral
filtering is applied in order to detect only photons which fall into the
non-distinguishable part of the spectrum. However, any form of spectral
filtering drastically reduces the count rate. This is especially
disadvantageous in multi-photon experiments where the coincidence
rate decreases rapidly with any filter loss. A scheme that
can eliminate the spectral distinguishability without significant loss of
signal will benefit these experiments greatly.
 One of the ways to overcome this problem is the spectral compensation scheme
proposed and first implemented by Kim \textit{et al.}~\cite{kim:02}.
In this report we will
present a detailed experimental study of the effectiveness of this method.

In the ``classic'' method of generating polarization-entangled photon pairs in
non-collinear type-II parametric down conversion, photon pairs are collected at
the intersection of the \textit{e} and \textit{o} polarized emission
cones~\cite{kwiat:95}. Their combined state covering polarization,
direction, and spectral fingerprint can be written as
\begin{equation}
\ket{\Psi} = {1\over\sqrt{2}}\left(
  \ket{H_e}_1\ket{V_o}_2\,+\,e^{i\delta}\ket{V_o}_1\ket{H_e}_2\right)\,,
\label{eq:gen_ouput_state_psi}
\end{equation}
where $\delta$ is the free phase between the states $\ket{H_e}_1\ket{V_o}_2$
and $\ket{V_o}_1\ket{H_e}_2$ corresponding to the two conversion paths.

\begin{figure}
\begin{center}
 \includegraphics[width=60mm]{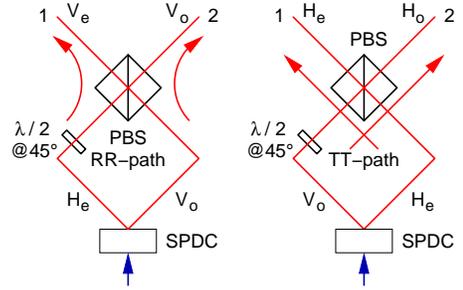}
\end{center}
\caption{(Color online) The possible paths of the photon pair generated in
  spontaneous parametric down conversion (SPDC) for the two corresponding
  down-converted components. The \textit{e} and \textit{o} polarized photons
  will exit at the different ports of the PBS independent of their
  polarization.}
\label{fig:pbs_comp}
\end{figure}

In the spectral compensation scheme (Fig.~\ref{fig:pbs_comp}), a half-wave
plate ($\lambda$/2) placed in one of the arm rotates the polarization by
$90^\circ$,  such that both photons arrive at the polarization beam splitter
(PBS) with the same polarization. The $\ket{H_e}_1\ket{V_o}_2$ combination is
transformed into $\ket{V_e}_1\ket{V_o}_2$, so both photons are
reflected by the PBS (RR path), while the
$\ket{V_o}_1\ket{H_e}_2$ combination is transformed into
$\ket{H_e}_1\ket{H_o}_2$, so both photons are transmitted
by the PBS (TT path). Regardless of their polarization state, photons carrying
the spectral fingerprint of \textit{o} and \textit{e}
polarization from the original conversion process will always emerge at a
corresponding port of the PBS. As long as there is no path difference between
the down conversion crystal and the PBS,
neither the arrival time nor the spectrum of the photon will reveal
information of the input polarization state, decoupling the temporal and
spectral degree of freedom from the polarization. The \textit{o} and
\textit{e} polarized photons need not arrive strictly simultaneously at the PBS
for the scheme to work, as shown in
various two-photon interference experiments~\cite{pittman:96, kim:03}.
Similarly to the Hong-Ou-Mandel interference of photon pairs~\cite{hong:87},
this scheme does not require path length stability to a fraction of the
wavelength, but only to a fraction of the coherence length of the photons. It
is also simple in the sense that it requires no special engineering of the
down-conversion medium or complex double-pass setups.
For a free phase $\delta=\pi$, the photon pairs are in the Bell state
\begin{equation}
\ket{\Phi^-} = {1\over\sqrt{2}}\left(\ket{H}_1\ket{H}_2 - \ket{V}_1\ket{V}_2
\right)\,,
\label{eq:gen_ouput_state_phi}
\end{equation}
which we will investigate for the rest of the paper.

\begin{figure}
\begin{center}
  \includegraphics[width=80mm]{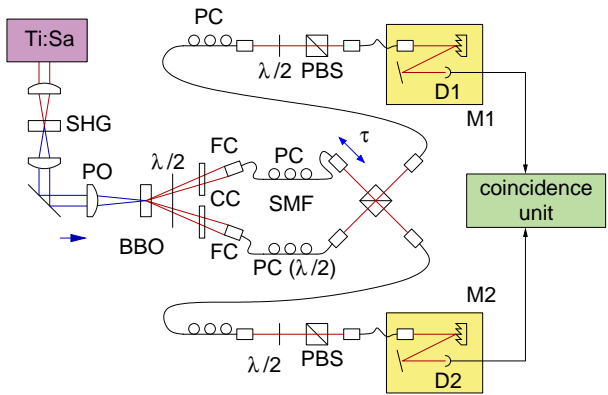}
\end{center}
\caption{(Color online) Experimental set-up. Photon pairs
  generated via SPDC in a nonlinear crystal (BBO) pumped by femtosecond
  optical pulses are collected into single-mode optical fibers (SMF). A
  half-wave plate ($\lambda$/2) and polarizing beam splitter (PBS) combination
  renders them spectrally indistinguishable. The down-converted photons then
  pass through polarization filters and subsequent grating monochromators for
  analysis.}
\label{fig:setup}
\end{figure}

We implemented a polarization-entangled photon pair source using type-II phase
matching in a crossed-ring configuration~\cite{kwiat:95} and use
polarization filters and grating monochromators to resolve the different
spectral components for both photons (Fig.~\ref{fig:setup}). 

The output of a Ti:Sapphire (Ti:Sa) laser with a repetition rate of
76\,MHz is frequency doubled (SHG) to pulses at
$\overline{\lambda}_p$ = 390\,nm with a spectral bandwidth of
$\Delta\lambda_p$\,$\approx$\,1.6\,nm full width at half maximum (FWHM). This
light beam (average power 900\,mW) is corrected for astigmatism and collimated
to a 
waist of 60\,${\mu}$m by the pump optics (PO). At the focus, a 2\,mm thick
Beta-Barium-Borate (BBO)
crystal cut for type-II phase matching with the extraordinary axis aligned to
the vertical polarization of the pump is used for down-conversion. A
combination of a $\lambda$/2 and a pair of compensation crystals (CC)
eliminates the temporal and transverse walk-off~\cite{kwiat:95}.  The free
phase $\delta$ is adjusted by tilting the CC. Down-converted photons are
collected into single mode optical fibers (SMF) with a spatial mode diameter matched to
that of the pump~\cite{kurtsiefer:01, bovino:03}. Polarization controllers (PC) ensure
that the SMF do not affect the polarization of the collected photons.

\begin{figure}
\begin{center}
  \includegraphics{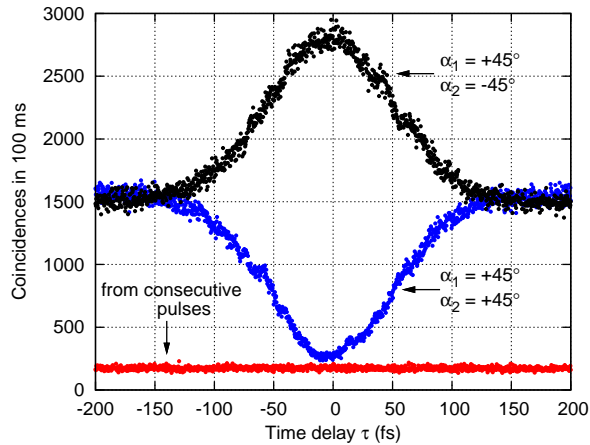}
\end{center}
\caption{(Color online) Polarization correlations measured in the
  $+45^\circ/-45^\circ$ basis as a function of delay $\tau$. Polarization
  analyzers were oriented at $\alpha_1$\,=\,$-\alpha_2$\,=\,$+45^\circ$ for
  the bump and at $\alpha_1$\,=\,$\alpha_2$\,=\,$+45^\circ$ for the dip. The
  bottom trace represents pair coincidences from consecutive pulses. Without
  correcting for higher order contribution, the visibility of the dip is
  85\,$\pm$\,2\,\%. The band of confidence for the corrected value is $[90\pm2\%, 96\pm3\%]$. Refer to the later text for details on the correction procedure.}
\label{fig:dip}
\end{figure}

In one of the spatial modes, the PC is adjusted such that the SMF rotates the
polarization by $90^\circ$. Both spatial modes
are then overlapped in a PBS. The relative
propagation delay $\tau$ to the PBS is adjusted by
varying the optical distance for one of the spatial modes. For $\tau=0$,
the two decay paths are rendered temporally and spectrally indistinguishable,
leaving the photon pair in a pure polarization state $\ket{\Phi^-}$. This behavior is shown in Fig.~\ref{fig:dip}, where the coincidences between photons exiting from the PBS are analyzed in an orthogonal basis as a function of the delay, showing a characteristic bump and dip at the point of maximal overlap.

Polarization analysis in each arm is performed by a combination of a
$\lambda/2$ and a PBS, allowing projections onto any arbitrary linear
polarization. We denote the direction of these linear polarizations by
their rotation $\alpha_1$ and $\alpha_2$ with respect to vertical. The
transmitted photons are transferred to grating monochromators (M1, M2) with
a resolution of 0.3\,nm FWHM, and detected with
passively quenched Silicon avalanche photodiodes (D1, D2). The detector
signals are analyzed for coincidences within a time window shorter
than the repetition period of the pump laser.

When sending the photons collected after the compensation scheme into
detectors D1 and D2, a coincidence rate of 105\,480\,s$^{-1}$ is observed. The
total coupling and detection efficiency (calculated from the ratio of pair
coincidences to single detector events in one arm) is 10.1\,\%.

To probe the quality of polarization entanglement between the photon pairs,
polarization correlations in two bases are measured. Conventionally, the
natural basis (i.e., the one aligned to the crystal axes) and a conjugate
basis are chosen, for our case the H/V and $+45/-45^\circ$ basis,
respectively. In the H/V basis, we expect to see nearly perfect
correlations due to the type-II conversion process. In the  $+45/-45^\circ$
basis, the level of (anti-)correlation will depend on the degree of indistinguishability
between the two decay paths. For $\ket{\Phi^-}$, perfect
anti-correlation in the $+45/-45^\circ$ basis is expected, but residual
distinguishability of the decay paths will deteriorate this.

To assess the degree of distinguishability, coincidences between the detectors
over a range of delays $\tau$ are recorded for
$\alpha_1$\,=\,$-\alpha_2$\,=\,$+45^\circ$. The result is shown in
Fig.~\ref{fig:dip}, which reveals clearly a bump for $\tau=0$. A fit to a
Gaussian distribution reveals a FWHM of approximately 100\,fs, corresponding to the coherence time of the down-converted photons. By choosing $\alpha_1=\alpha_2=+45^\circ$, a corresponding dip in
coincidences is observed. From Fig.~\ref{fig:dip}, the
maximal visibility of the dip is
85\,$\pm$\,2\,\%.

\begin{figure}
\begin{center}
  \includegraphics{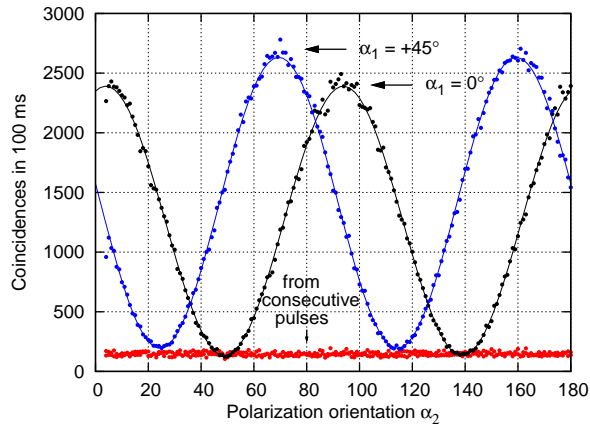}
\end{center}
\caption{(Color online) Polarization correlations in the H/V and
  $+45^\circ/-45^\circ$ bases. The bottom trace represents pair coincidences
  from consecutive pulses. Without correcting for any higher order contribution,
  we observed direct visibilities of $V_{HV}$\,=\,90.0\,$\pm$\,0.4\% and
  $V_{45}$\,=\,86.8\,$\pm$\,0.4\%.} \label{fig:visibility}
\end{figure}

The polarization entanglement of the photon pairs was characterized  by
measuring the visibilities $V_{HV}$ and $V_{45}$ in the H/V and
$+45^\circ$/-$45^\circ$ basis, respectively. We obtain $V_{HV}$ ($V_{45}$) by
fixing $\alpha_1$ at $0^\circ$ ($+45^\circ$), and rotating the
orientation of the other analyzer while recording the coincidences.
Without spectral filtering, we obtain results $V_{HV}$\,=\,90.0\,$\pm$\,0.4\%
and $V_{45}$\,=\,86.8\,$\pm$\,0.4\% (see Fig.~\ref{fig:visibility}).

 However, due to the high instantaneous power involved in the
 femtosecond-pumped down-conversion, higher order processes (mainly
 four-photon generation) become significant, and it is important to
 quantify their contribution. When observing only two-fold coincidences, this
 four-photon contribution will lead to uncorrelated events lowering the
 two-photon visibilities. To estimate this four-photon contribution,
 we record coincidences between consecutive pulses in the same
 run. Following an argument put forward in~\cite{riedmatten:04}, the
 coincidence rate between consecutive pulses is the same as the rate of
 distinguishable pairs generated in the same pulse. If the two photon pairs are
 indistinguishable, the four-photon contribution to the two-photon coincidence rate will be half of the
 pair coincidence rate between consecutive pulses. This allows us to come up
 with a
 lower and upper bound for the four-photon generation rate in the
 setup. Correcting for this higher order contribution, we obtain bands of
 confidence for the visibilities, $V_{HV}\in\,[95.1\pm0.5\%, 100.8\pm0.5\%]$
 and $V_{45}\in\,[92.0\pm0.4\%,97.9\pm0.5\%]$ for the process leading to pairs
 only.

 Both the bounds for $V_{45}$ are significantly higher than the results obtained without spectral compensation~\cite{poh:07}, where we see
 $V_{45}\in[69.1\pm0.3\%, 72.9\pm0.3\%]$ without spectral filtering, and
 $V_{45}\in[83.1\pm0.3\%, 85.9\pm0.3\%]$ with spectral filtering. This shows
 that the spectral compensation scheme has eliminated the distinguishability
 between the two down-converted components.

\begin{figure}
\begin{center}
  \includegraphics{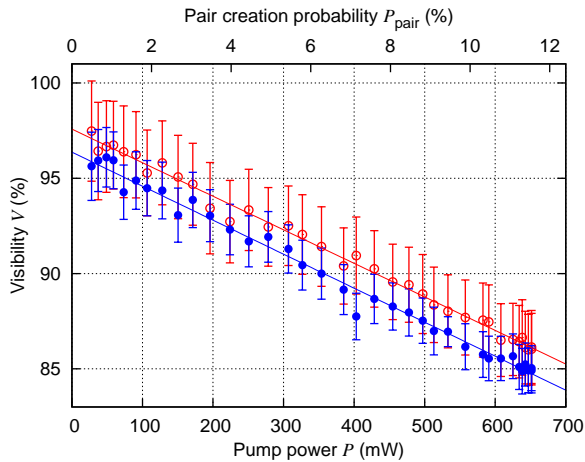}
\end{center}
\caption{(Color online) Visibility $V_{HV}$ (open circles) and $V_{45}$ (solid
  circles) measured as a function of the pump power. The probability of
  creating a pair $P_{\mathrm{pair}}$ (top axis) is proportional to the pump
  power. Solid lines show linear fits to the visibility reduction. From the
  slope, a pair generation probability can be derived via
  Eq.~\ref{eq:vis_pwr_vary} (top axis). 
  At low power, the coincidences are dominated by the
  contribution from first order down-conversion. The extrapolated
  visibilities at $P=0$ $V_{HV}=97.6\pm0.1\%$ and $V_{45}=96.4\pm0.1\%$.}
\label{fig:vis_pwr}
\end{figure}

To provide a consistency check for the correction procedure, measurements of
the visibilities are made with various pump powers. A model describing the
dependence of visibility on pump power is described in~\cite{marcikic:02}. It
assumes that the detected pair rate has a contribution $R_2$ from pairs
generated in the same birth process, and a contribution
$R_4$ from partially detected, incoherent double pair events. They can be
written as
\begin{eqnarray}
R_2&=&P_{\mathrm{pair}}\,\frac{1+\cos\theta}{2},\nonumber \\
R_4&=&4\,P_{\mathrm{double\;pair}}\,\frac{2+\cos\theta}{4}\,,
\end{eqnarray}
where $\theta=\alpha_1-\alpha_2$, and $P_{\mathrm{pair}}$ and
$P_{\mathrm{double\;pair}}$ are the probabilities for creating a pair and an
incoherent double pair per pulse, respectively.
The first one can be written as
\begin{equation}
P_{\mathrm{pair}}=\frac{S}{\eta_c\eta_qf}\,,
\label{eq:ppair}
\end{equation}
where $S$ is the rate of detector events on one side, $\eta_c$ characterizes
the coupling efficiency, $\eta_q$ is the quantum efficiency of the
detectors, and $f$ the repetition rate of the laser.
Assuming a Poissonian
distribution in the counting of incoherent pairs events, one finds
$P_{\mathrm{double\;pair}}=P^2_{\mathrm{pair}}/2$. The visibility $V$
obtained from the sum of $R_2$ and $R_4$ is
\begin{equation}
V\approx V_{\mathrm{max}}-P_{\mathrm{pair}}\,,
\label{eq:vis_pwr_vary}
\end{equation}
with $V_{\mathrm{max}}$ being the limit for the visibility at low pump
power.
 Figure~\ref{fig:vis_pwr} shows the visibilities observed for different pump
 powers, exhibiting a linear decrease with power as expected
 according to  Eq.~\ref{eq:vis_pwr_vary}. The slope of both visibility
 measurements coincide ($0.0177\pm0.0003$\%\,mW$^{-1}$) and allows to rescale
 power  into pair probability (see top axis on figure). 
 From the pair probability and a single detector event rate (corrected for
 saturation effects), a combined detector/coupling efficiency of 11.3\% can be
 derived via Eq.~(\ref{eq:ppair}). 

The limit $V_{\mathrm{max}}$ for
 the visibility at low pump power are $V_{HV}=97.6\pm0.1\%$ and $V_{45}=96.4\pm0.1\%$ in
agreement with results from the correction procedure.

To understand the joint spectral properties of the polarization correlations,
we measured the joint spectrum of the photon pairs generated from each of the
two decay paths. This is done by fixing the polarization analyzers to the
natural basis of the down-conversion crystal, selecting either the $H_1H_2$ or
$V_1V_2$ decay path. The spectra are taken
with a resolution of 0.5\,nm and an integration time of 30\,s for each
wavelength pair.

\begin{figure}
\begin{center}
\includegraphics[width=80mm]{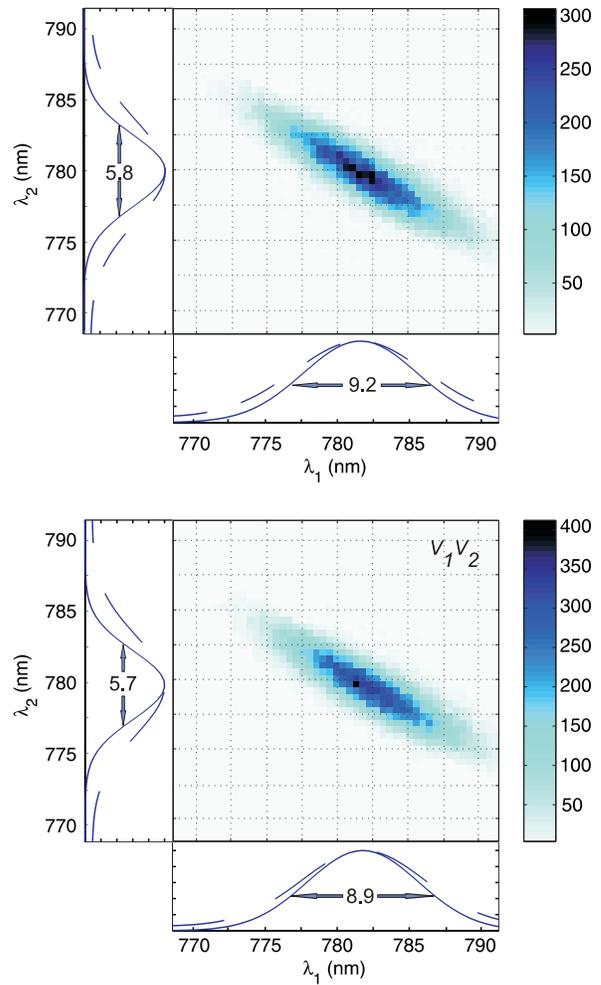}
\end{center}
\caption{(Color online) Joint spectra of coincidence counts in 30\,s for $H_1H_2$
  (upper panel) and $V_1V_2$ (lower panel) polarizations. The spectra
  corresponding to the two decay paths $RR$ and $TT$ are almost identical with
  the exception of the difference in the maximum count
  rate recorded. Differences between the widths of the marginal (solid trace)
  and the single photon spectra (dashed traces), as well as between the
  \textit{e} and \textit{o} polarization are observed as expected.}
\label{fig:hv_vh_joint}
\end{figure}

\begin{figure}
\begin{center}
\includegraphics[width=80mm]{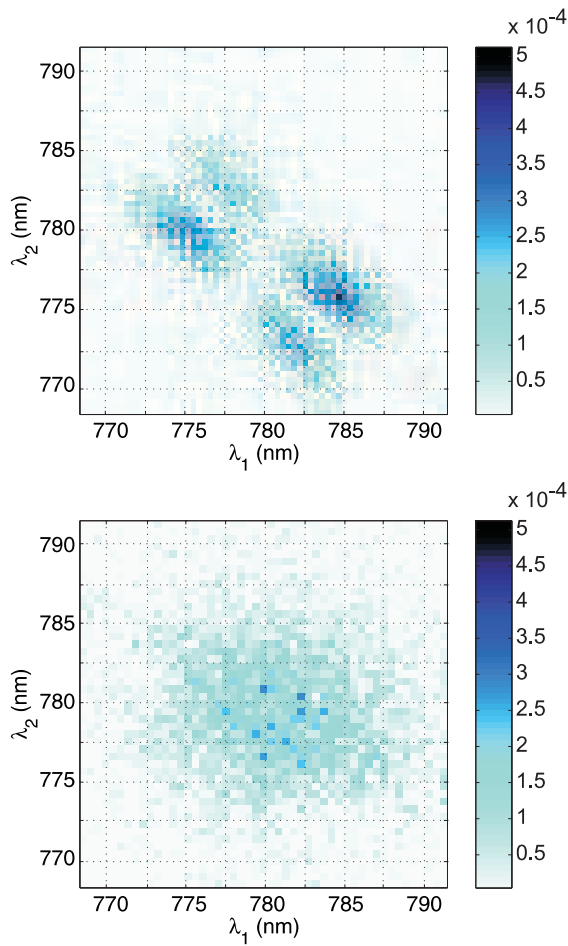}
\end{center}
\caption{(Color online) Joint spectra of coincidences measured for the
  $+45^\circ$/$+45^\circ$ polarization combination. The counts are normalized
  to the total events in the spectral mapping experiments for both
  configurations. Without the spectral compensation scheme
  (upper panel), the joint spectrum exhibits four regions of higher count
  rate, comprising a fraction of 0.14 of all events. They correspond to areas
  with an imbalance of the two decay paths. With the spectral compensation
  scheme (lower panel), the joint spectrum, a fraction of 0.10 of
  the total events, exhibits a distribution of
  uncorrelated pair events about the degenerate wavelength which is compatible
  with estimates of the four-photon contribution.}
\label{fig:45joint}
\end{figure}

Our results (shown in Fig.~\ref{fig:hv_vh_joint})
have nearly identical distributions, with the exception of their maximum
coincidence rate. A balanced contribution between the $H_1H_2$ and $V_1V_2$
decay path is found at all wavelength pairs, compatible with state
$\ket{\Phi^-}$ at every point. This is in contrast with results obtained
without spectral
compensation~\cite{poh:07}, with the different spectral fingerprints of the
two decay paths.

The marginal distributions $\lambda_1$, $\lambda_2$ exhibit widths of
$\Delta\lambda_{m1}$\,=\,9.2\,$\pm$\,0.3\,nm,
$\Delta\lambda_{m2}$\,=\,5.8\,$\pm$\,0.2\,nm  (FWHM) for the $H_1H_2$
combination. A comparable distribution is also observed for the $V_1V_2$ joint
spectrum (Fig.~\ref{fig:hv_vh_joint}, lower panel).
Thus, the spectral degree of
freedom no longer reveals any information on the corresponding polarization
state.

The spectral distribution of the photons collected in each spatial mode
obtained from the same run have central wavelengths of
$\overline{\lambda}_H=781.55\pm0.03$\,nm and
$\overline{\lambda}_V=780.19\pm0.01$\,nm, and a FWHM of
$\Delta\lambda_H=10.7\pm0.1$\,nm and
$\Delta\lambda_V=8.30\pm0.05$\,nm, respectively. We note that the single
photon spectral distributions (dashed lines in Fig.~\ref{fig:hv_vh_joint}) are
wider than the marginals, $\Delta\lambda_H=8.3$\,nm,
$\Delta\lambda_V=10.4$\,nm. This disparity is caused by the difference in the width of the spectral
distribution between the \textit{e} and \textit{o} polarized photon. In the
SPDC process, the spectral distribution of down-converted photons is connected
to their angular dispersion. For a certain acceptance
angle subtended by the collection, the \textit{e} polarised photons (which have
a narrower spread in the emission direction) will be collected more efficiently
than the \textit{o} polarized photon. Thus, not every photon detected in one
spatial mode has its twin in the other mode~\cite{poh:07}.

The joint spectra for polarizations in the complementary basis,
($+45^\circ$/$+45^\circ$), are shown in Fig.~\ref{fig:45joint}, normalized to
the total events both in the $+45^\circ/+45^\circ$  and
$+45^\circ/-45^\circ$ basis. The upper panel shows the result without spectral
compensation~\cite{poh:07} for comparison. It reveals regions with higher
rates, indicating 
an imbalance between the two down-converted components in those areas. With
the spectral compensation (lower panel), the distribution of uncorrelated
pair events is compatible with the four-photon contribution. This can be seen from the marginal distribution of the joint spectrum, $\Delta\lambda_{m1}$\,=\,10.8\,$\pm$\,0.4\,nm, $\Delta\lambda_{m2}$\,=\,8.4\,$\pm$\,0.2\,nm (FWHM) which is identical to the distribution of the photon collected in each spatial mode.

To characterize the distinguishability between the two decay paths, we also
need to look at the corresponding $+45^\circ$/$-45^\circ$ joint
spectrum. Together with this measurement, the visibility $V_{45}$ can be
reconstructed by summing over all wavelengths:
\begin{equation}
  V_{45}=\frac{\sum\limits_{\lambda_1, \lambda_2}c^{+,-}(\lambda_1, \lambda_2)
    - \sum\limits_{\lambda_1, \lambda_2}c^{+,+}(\lambda_1, \lambda_2)}
  {\sum\limits_{\lambda_1,  \lambda_2}c^{+,-}(\lambda_1, \lambda_2) + 
    \sum\limits_{\lambda_1, \lambda_2}c^{+,+}(\lambda_1, \lambda_2)}\,,
\label{eq:visibility_joint_spectrum}
\end{equation}
where the $c^{+,\pm}(\lambda_1, \lambda_2)$ are coincidence events detected for
various wavelength pairs, and +-- and ++ refers to the $+45^\circ$/$-45^\circ$
and $+45^\circ$/$+45^\circ$ polarizer settings.

After correcting for the four-photon contribution, we get
$V_{45}\in\,$[73.8\,$\pm$\,0.5\%, 80.2\,$\pm$\,0.6\%] without spectral compensation. With spectral compensation, we get $V_{45}\in\,$[89.4\,$\pm$\,0.5\%, 100.4\,$\pm$\,0.6\%].

In summary, the compensation scheme eliminated the spectral distinguishability between the two decay paths. This is demonstrated by the identical joint spectra measured in the natural basis of the down-conversion and direct correlation measurements at different power levels. The balanced contribution between the two down-conversion paths for all spectral components does not
reveal any information about the polarization state, thus entanglement quality
is preserved when the spectral degree of freedom is ignored. After taking the
higher order contributions into consideration, we achieved a high visibility
of $V_{45}=97.9\pm0.5$\% in the complementary basis without the need
of spectral filtering. The simplicity and effectiveness of this scheme make it
a useful addition to the toolkit of techniques used for efficiently preparing
entangled states of two and more photons.

This work is supported by the National Research Foundation \& Ministry of
Education, Singapore, and ASTAR under SERC grant No.~052\,101\,0043.

\bibliographystyle{apsrev}

\begin{thebibliography}{26}
\expandafter\ifx\csname
natexlab\endcsname\relax\def\natexlab#1{#1}\fi
\expandafter\ifx\csname bibnamefont\endcsname\relax
  \def\bibnamefont#1{#1}\fi
\expandafter\ifx\csname bibfnamefont\endcsname\relax
  \def\bibfnamefont#1{#1}\fi
\expandafter\ifx\csname citenamefont\endcsname\relax
  \def\citenamefont#1{#1}\fi

\bibitem{kim:02} Y.-H. Kim and W.~P. Grice, J. Mod. Opt. \textbf{49},
2309 (2002)

\bibitem[{\citenamefont{Bouwmeester et~al.}(2001)\citenamefont{Bouwmeester, Ekert, and Zeilinger}}]{bouwmeester:01}
\bibinfo{author}{\bibfnamefont{D.}~\bibnamefont{Bouwmeester}},
\bibinfo{author}{\bibfnamefont{A.}~\bibnamefont{Ekert}}, \bibnamefont{and}
\bibinfo{author}{\bibfnamefont{A.}~\bibnamefont{Zeilinger}},
\emph{\bibinfo{title}{The physics of quantum information}}
(\bibinfo{publisher}{Springer}, \bibinfo{year}{2001}).

\bibitem[{\citenamefont{Burnham and Weinberg}(1970)}]{burnham:70}
\bibinfo{author}{\bibfnamefont{D.~C.} \bibnamefont{Burnham}} \bibnamefont{and}
\bibinfo{author}{\bibfnamefont{D.~L.} \bibnamefont{Weinberg}},
\bibinfo{journal}{Phys. Rev. Lett.} \textbf{\bibinfo{volume}{25}},
\bibinfo{pages}{84} (\bibinfo{year}{1970}).

\bibitem[{\citenamefont{Klyshko}(1989)}]{klyshko:89}
\bibinfo{author}{\bibfnamefont{D.~N.} \bibnamefont{Klyshko}},
\emph{\bibinfo{title}{Photons and Nonlinear Optics}}
(\bibinfo{publisher}{Gordon and Breach Science Publishers},
\bibinfo{address}{New York}, \bibinfo{year}{1989}).

\bibitem[{\citenamefont{Kwiat et~al.}(1995)\citenamefont{Kwiat, Mattle, Weinfurter, Zeilinger, Sergienko, and Shih}}]{kwiat:95}
\bibinfo{author}{\bibfnamefont{P.~G.} \bibnamefont{Kwiat}},
\bibinfo{author}{\bibfnamefont{K.}~\bibnamefont{Mattle}},
\bibinfo{author}{\bibfnamefont{H.}~\bibnamefont{Weinfurter}},
\bibinfo{author}{\bibfnamefont{A.}~\bibnamefont{Zeilinger}},
\bibinfo{author}{\bibfnamefont{A.~V.} \bibnamefont{Sergienko}},
\bibnamefont{and} \bibinfo{author}{\bibfnamefont{Y.}~\bibnamefont{Shih}},
\bibinfo{journal}{Phys. Rev. Lett.} \textbf{\bibinfo{volume}{75}},
\bibinfo{pages}{4337} (\bibinfo{year}{1995}).

\bibitem[{\citenamefont{Brendel et~al.}(1992)\citenamefont{Brendel, Mohler, and Martienssen}}]{brendel:92}
\bibinfo{author}{\bibfnamefont{J.}~\bibnamefont{Brendel}},
\bibinfo{author}{\bibfnamefont{E.}~\bibnamefont{Mohler}}, \bibnamefont{and}
\bibinfo{author}{\bibfnamefont{W.}~\bibnamefont{Martienssen}},
\bibinfo{journal}{Europhys. Lett.} \textbf{\bibinfo{volume}{20}},
\bibinfo{pages}{575} (\bibinfo{year}{1992}).

\bibitem[{\citenamefont{Kwiat et~al.}(1999)\citenamefont{Kwiat, Waks, White, Appelbaum, and Eberhard}}]{kwiat:99}
\bibinfo{author}{\bibfnamefont{P.~G.}~\bibnamefont{Kwiat}},
\bibinfo{author}{\bibfnamefont{E.}~\bibnamefont{Waks}},
\bibinfo{author}{\bibfnamefont{A.~G.}~\bibnamefont{White}},
\bibinfo{author}{\bibfnamefont{I.}~\bibnamefont{Appelbaum}},
\bibnamefont{and} \bibinfo{author}{\bibfnamefont{P.~H.}
\bibnamefont{Eberhard}}, \bibinfo{journal}{Phys. Rev. A}
\textbf{\bibinfo{volume}{60}}, \bibinfo{pages}{R773} (\bibinfo{year}{1999}).

\bibitem[{\citenamefont{Jennewein et~al.}(2000)\citenamefont{Jennewein, Simon, Weihs, Weinfurter, and Zeilinger}}]{jennewein:00}
\bibinfo{author}{\bibfnamefont{T.}~\bibnamefont{Jennewein}},
\bibinfo{author}{\bibfnamefont{C.}~\bibnamefont{Simon}},
\bibinfo{author}{\bibfnamefont{G.}~\bibnamefont{Weihs}},
\bibinfo{author}{\bibfnamefont{H.}~\bibnamefont{Weinfurter}}, \bibnamefont{and}
\bibinfo{author}{\bibfnamefont{A.}~\bibnamefont{Zeilinger}},
\bibinfo{journal}{Phys. Rev. Lett.} \textbf{\bibinfo{volume}{84}},
\bibinfo{pages}{4729} (\bibinfo{year}{2000}).

\bibitem[{\citenamefont{Gr\"{o}blacher et~al.}(2007)\citenamefont{Gr\"{o}blacher, Paterek, Kaltenbaek, Brukner, \.{Z}ukowski, Aspelmeyer, and Zeilinger}}]{groblacher:07}
\bibinfo{author}{\bibfnamefont{S.}~\bibnamefont{Gr\"{o}blacher}},
\bibinfo{author}{\bibfnamefont{T.}~\bibnamefont{Paterek}},
\bibinfo{author}{\bibfnamefont{R.}~\bibnamefont{Kaltenbaek}},
\bibinfo{author}{\bibfnamefont{\v{C}.}~\bibnamefont{Brukner}},
\bibinfo{author}{\bibfnamefont{M.}~\bibnamefont{\.{Z}ukowski}},
\bibinfo{author}{\bibfnamefont{M.}~\bibnamefont{Aspelmeyer}}, \bibnamefont{and}
\bibinfo{author}{\bibfnamefont{A.}~\bibnamefont{Zeilinger}},
\bibinfo{journal}{Nature} \textbf{\bibinfo{volume}{446}},
\bibinfo{pages}{871} (\bibinfo{year}{2007}).

\bibitem[{\citenamefont{Branciard et~al.}(2008)\citenamefont{Branciard, Brunner, Gisin, Kurtsiefer, Lamas-Linares, Ling, and Scarani}}]{branciard:08}
\bibinfo{author}{\bibfnamefont{C.}~\bibnamefont{Branciard}},
\bibinfo{author}{\bibfnamefont{N.}~\bibnamefont{Brunner}},
\bibinfo{author}{\bibfnamefont{N.}~\bibnamefont{Gisin}},
\bibinfo{author}{\bibfnamefont{C.}~\bibnamefont{Kurtsiefer}},
\bibinfo{author}{\bibfnamefont{A.}~\bibnamefont{Lamas-Linares}},
\bibinfo{author}{\bibfnamefont{A.}~\bibnamefont{Ling}}, \bibnamefont{and}
\bibinfo{author}{\bibfnamefont{V.}~\bibnamefont{Scarani}},
\bibinfo{journal}{Nature Phys.} \textbf{\bibinfo{volume}{4}},
\bibinfo{pages}{681} (\bibinfo{year}{2008}).

\bibitem[{\citenamefont{Bouwmeester et~al}(1997)}]{bouwmeester:97}
\bibinfo{author}{\bibfnamefont{D.} \bibnamefont{Bouwmeester}},
\bibinfo{author}{\bibfnamefont{J.-W.} \bibnamefont{Pan}},
\bibinfo{author}{\bibfnamefont{K.} \bibnamefont{Mattle}},
\bibinfo{author}{\bibfnamefont{M.} \bibnamefont{Eibl}},
\bibinfo{author}{\bibfnamefont{H.} \bibnamefont{Weinfurter}}, \bibnamefont{and}
\bibinfo{author}{\bibfnamefont{A.} \bibnamefont{Zeilinger}},
\bibinfo{journal}{Nature (London)} \textbf{\bibinfo{volume}{390}},
\bibinfo{pages}{575} (\bibinfo{year}{1997}).

\bibitem[{\citenamefont{Linares et~al}(2002)}]{linares:02}
\bibinfo{author}{\bibfnamefont{A.} \bibnamefont{Lamas-Linares}},
\bibinfo{author}{\bibfnamefont{J.C.} \bibnamefont{Howell}},
\bibinfo{author}{\bibfnamefont{C.} \bibnamefont{Simon}}, \bibnamefont{and}
\bibinfo{author}{\bibfnamefont{D.} \bibnamefont{Bouwmeester}},
\bibinfo{journal}{Science} \textbf{\bibinfo{volume}{296}},
\bibinfo{pages}{712} (\bibinfo{year}{2002}).

\bibitem[{\citenamefont{Gaertner et~al}(2003)}]{gaertner:03}
\bibinfo{author}{\bibfnamefont{S.} \bibnamefont{Gaertner}},
\bibinfo{author}{\bibfnamefont{M.} \bibnamefont{Bourennane}},
\bibinfo{author}{\bibfnamefont{M.} \bibnamefont{Eibl}},
\bibinfo{author}{\bibfnamefont{C.} \bibnamefont{Kurtsiefer}}, \bibnamefont{and}
\bibinfo{author}{\bibfnamefont{H.} \bibnamefont{Weinfurter}},
\bibinfo{journal}{Appl. Phys. B} \textbf{\bibinfo{volume}{77}},
\bibinfo{pages}{803} (\bibinfo{year}{2003}).

\bibitem[{\citenamefont{Goebel et~al.}(2008)}]{goebel:03}
\bibinfo{author}{\bibfnamefont{A.~M.} \bibnamefont{Goebel}},
\bibinfo{author}{\bibfnamefont{C.} \bibnamefont{Wagenknecht}},
\bibinfo{author}{\bibfnamefont{Q.} \bibnamefont{Zhang}},
\bibinfo{author}{\bibfnamefont{Y.-A.} \bibnamefont{Chen}},
\bibinfo{author}{\bibfnamefont{K.} \bibnamefont{Chen}},
\bibinfo{author}{\bibfnamefont{J.} \bibnamefont{Schmiedmayer}}, \bibnamefont{and}
\bibinfo{author}{\bibfnamefont{J.-W.} \bibnamefont{Pan}},
\bibinfo{journal}{Phys. Rev. Lett.} \textbf{\bibinfo{volume}{101}},
\bibinfo{pages}{080403} (\bibinfo{year}{2008}).

\bibitem[{\citenamefont{Keller and Rubin}(1997)}]{keller:97}
\bibinfo{author}{\bibfnamefont{T.~E.} \bibnamefont{Keller}} \bibnamefont{and}
\bibinfo{author}{\bibfnamefont{M.~H.} \bibnamefont{Rubin}},
\bibinfo{journal}{Phys. Rev. A} \textbf{\bibinfo{volume}{56}},
\bibinfo{pages}{1534} (\bibinfo{year}{1997}).

\bibitem[{\citenamefont{Grice and Walmsley}(1997)}]{grice:97}
\bibinfo{author}{\bibfnamefont{W.~P.} \bibnamefont{Grice}} \bibnamefont{and}
\bibinfo{author}{\bibfnamefont{I.~A.} \bibnamefont{Walmsley}},
\bibinfo{journal}{Phys. Rev. A} \textbf{\bibinfo{volume}{56}},
\bibinfo{pages}{1627} (\bibinfo{year}{1997}).

\bibitem[{\citenamefont{Grice et~al}(1998)}]{grice:98}
\bibinfo{author}{\bibfnamefont{W.~P.} \bibnamefont{Grice}},
\bibinfo{author}{\bibfnamefont{R.} \bibnamefont{Erdmann}},
\bibinfo{author}{\bibfnamefont{I.~A.} \bibnamefont{Walmsley}}, \bibnamefont{and}
\bibinfo{author}{\bibfnamefont{D.} \bibnamefont{Branning}},
\bibinfo{journal}{Phys. Rev. A} \textbf{\bibinfo{volume}{57}},
\bibinfo{pages}{R2289} (\bibinfo{year}{1998}).

\bibitem[{\citenamefont{Atat\"{u}re et~al}(1999)}]{atature:99}
\bibinfo{author}{\bibfnamefont{M.} \bibnamefont{Atat\"{u}re}},
\bibinfo{author}{\bibfnamefont{A.~V.} \bibnamefont{Sergienko}} ,
\bibinfo{author}{\bibfnamefont{B.~M.} \bibnamefont{Jost}} ,
\bibinfo{author}{\bibfnamefont{B.~E.~S.} \bibnamefont{Saleh}}, \bibnamefont{and}
\bibinfo{author}{\bibfnamefont{M.~C.} \bibnamefont{Teich}},
\bibinfo{journal}{Phys. Rev. Lett.} \textbf{\bibinfo{volume}{83}},
\bibinfo{pages}{1323} (\bibinfo{year}{1999}).

\bibitem[{\citenamefont{Kim et~al}(2005)}]{kim:05}
\bibinfo{author}{\bibfnamefont{Y.-H.} \bibnamefont{Kim}} \bibnamefont{and}
\bibinfo{author}{\bibfnamefont{W.~P.} \bibnamefont{Grice}},
\bibinfo{journal}{Opt. Lett.} \textbf{\bibinfo{volume}{30}},
\bibinfo{pages}{908} (\bibinfo{year}{2005}).

\bibitem[{\citenamefont{Wasilewski et~al}(2006)}]{wasilewski:06}
\bibinfo{author}{\bibfnamefont{W.} \bibnamefont{Wasilewski}},
\bibinfo{author}{\bibfnamefont{P.} \bibnamefont{Wasylczyk}} ,
\bibinfo{author}{\bibfnamefont{P.} \bibnamefont{Kolenderski}} ,
\bibinfo{author}{\bibfnamefont{K.} \bibnamefont{Banaszek}}, \bibnamefont{and}
\bibinfo{author}{\bibfnamefont{C.} \bibnamefont{Radzewicz}},
\bibinfo{journal}{Opt. Lett.} \textbf{\bibinfo{volume}{31}},
\bibinfo{pages}{1130} (\bibinfo{year}{2006}).

\bibitem[{\citenamefont{Poh et~al.}(2007)\citenamefont{Poh, Lum, Marcikic, Lamas-Linares, and Kurtsiefer}}]{poh:07}
\bibinfo{author}{\bibfnamefont{H.~S.}~\bibnamefont{Poh}},
\bibinfo{author}{\bibfnamefont{C.~Y.}~\bibnamefont{Lum}},
\bibinfo{author}{\bibfnamefont{I.}~\bibnamefont{Marcikic}},
\bibinfo{author}{\bibfnamefont{A.}~\bibnamefont{Lamas-Linares}}, \bibnamefont{and}
\bibinfo{author}{\bibfnamefont{C.}~\bibnamefont{Kurtsiefer}},
\bibinfo{journal}{Phys. Rev. A} \textbf{\bibinfo{volume}{75}},
\bibinfo{pages}{043816} (\bibinfo{year}{2007}).

\bibitem[{\citenamefont{Avenhaus et~al.}(2009)}]{avenhaus:09}
\bibinfo{author}{\bibfnamefont{M.} \bibnamefont{Avenhaus}},
\bibinfo{author}{\bibfnamefont{M.~V.} \bibnamefont{Chekhova}},
\bibinfo{author}{\bibfnamefont{L.~A.} \bibnamefont{Leuchs}}, \bibnamefont{and}
\bibinfo{author}{\bibfnamefont{C.} \bibnamefont{Silberhon}},
\bibinfo{journal}{Phys. Rev. A.} \textbf{\bibinfo{volume}{79}},
\bibinfo{pages}{043836} (\bibinfo{year}{2009}).

\bibitem[{\citenamefont{Erdmann et~al.}(2000)}]{erdmann:00}
\bibinfo{author}{\bibfnamefont{R.} \bibnamefont{Erdmann}},
\bibinfo{author}{\bibfnamefont{D.} \bibnamefont{Branning}},
\bibinfo{author}{\bibfnamefont{W.~P.} \bibnamefont{Grice}},  \bibnamefont{and}
\bibinfo{author}{\bibfnamefont{I.~A.} \bibnamefont{Walmsley}},
\bibinfo{journal}{Phys. Rev. A} \textbf{\bibinfo{volume}{62}},
\bibinfo{pages}{053810} (\bibinfo{year}{2000}).

\bibitem[{\citenamefont{Hodelin et~al.}(2006)}]{hodelin:06}
\bibinfo{author}{\bibfnamefont{J.~F.} \bibnamefont{Hodelin}},
\bibinfo{author}{\bibfnamefont{G.} \bibnamefont{Khoury}}, \bibnamefont{and}
\bibinfo{author}{\bibfnamefont{D.} \bibnamefont{Bouwmeester}},
\bibinfo{journal}{Phys. Rev. A} \textbf{\bibinfo{volume}{74}},
\bibinfo{pages}{013802} (\bibinfo{year}{2006}).

\bibitem[{\citenamefont{Branning}(1999)\citenamefont{Branning}}]{branning:99}
\bibinfo{author}{\bibfnamefont{D.}~\bibnamefont{Branning}},
\bibinfo{author}{\bibfnamefont{W.~P.} \bibnamefont{Grice}},
\bibinfo{author}{\bibfnamefont{R.} \bibnamefont{Erdmann}}, \bibnamefont{and}
\bibinfo{author}{\bibfnamefont{I.~A.} \bibnamefont{Walmsley}},
\bibinfo{journal}{Phys. Rev. Lett.} \textbf{\bibinfo{volume}{83}},
\bibinfo{pages}{955} (\bibinfo{year}{1999}).

\bibitem[{\citenamefont{Pittman et~al.}(1996)\citenamefont{Pittman, Strekalov Migdall Rubin, Sergienko, and Shih}}]{pittman:96}
\bibinfo{author}{\bibfnamefont{T.~B.} \bibnamefont{Pittman}},
\bibinfo{author}{\bibfnamefont{D.~V.}~\bibnamefont{Strekalov}},
\bibinfo{author}{\bibfnamefont{A.}~\bibnamefont{Migdall}},
\bibinfo{author}{\bibfnamefont{M.~H}~\bibnamefont{Rubin}},
\bibinfo{author}{\bibfnamefont{A.~V.} \bibnamefont{Sergienko}},
\bibnamefont{and} \bibinfo{author}{\bibfnamefont{Y.}~\bibnamefont{Shih}},
\bibinfo{journal}{Phys. Rev. Lett.} \textbf{\bibinfo{volume}{77}},
\bibinfo{pages}{1917} (\bibinfo{year}{1996}).

\bibitem[{\citenamefont{Kim}(2003)\citenamefont{Kim}}]{kim:03}
\bibinfo{author}{\bibfnamefont{Y.-H.}~\bibnamefont{Kim}},
\bibinfo{journal}{Phys. Lett. A} \textbf{\bibinfo{volume}{315}},
\bibinfo{pages}{352} (\bibinfo{year}{2003}).

\bibitem[{\citenamefont{Hong et~al.}(1987)\citenamefont{Hong, Ou, and Mandel}}]{hong:87}
\bibinfo{author}{\bibfnamefont{C.~K.}~\bibnamefont{Hong}},
\bibinfo{author}{\bibfnamefont{Z.~Y.}~\bibnamefont{Ou}}, \bibnamefont{and}
\bibinfo{author}{\bibfnamefont{L.}~\bibnamefont{Mandel}},
\bibinfo{journal}{Phys. Rev. Lett.} \textbf{\bibinfo{volume}{59}},
\bibinfo{pages}{2044} (\bibinfo{year}{1987}).

\bibitem[{\citenamefont{Kursiefer et~al.}(2001)\citenamefont{Kursiefer, Oberparleiter, and Weinfurter}}]{kurtsiefer:01}
\bibinfo{author}{\bibfnamefont{C.}~\bibnamefont{Kurtsiefer}},
\bibinfo{author}{\bibfnamefont{M.}~\bibnamefont{Oberparleiter}}, \bibnamefont{and}
\bibinfo{author}{\bibfnamefont{H.}~\bibnamefont{Weinfurter}},
\bibinfo{journal}{Phys. Rev. A} \textbf{\bibinfo{volume}{64}},
\bibinfo{pages}{023802} (\bibinfo{year}{2001}).

\bibitem[{\citenamefont{Bovino et~al.}(2003)\citenamefont{Bovino, Varisco, Colla, Castagnoli, Giuseppe, and Sergienko}}]{bovino:03}
\bibinfo{author}{\bibfnamefont{F.~A.}~\bibnamefont{Bovino}},
\bibinfo{author}{\bibfnamefont{P.}~\bibnamefont{Varisco}},
\bibinfo{author}{\bibfnamefont{A.~M.}~\bibnamefont{Colla}},
\bibinfo{author}{\bibfnamefont{G.}~\bibnamefont{Castagnoli}},
\bibinfo{author}{\bibfnamefont{G.~D.}~\bibnamefont{Giuseppe}}, \bibnamefont{and}
\bibinfo{author}{\bibfnamefont{A.~V.}~\bibnamefont{Sergienko}},
\bibinfo{journal}{Opt. Commun.} \textbf{\bibinfo{volume}{227}},
\bibinfo{pages}{343} (\bibinfo{year}{2003}).

\bibitem{riedmatten:04}
\bibinfo{author}{\bibfnamefont{H.}~\bibnamefont{de Riedmatten}},
\bibinfo{author}{\bibfnamefont{V.}~\bibnamefont{Scarani}},
\bibinfo{author}{\bibfnamefont{I.}~\bibnamefont{Marcikic}},
\bibinfo{author}{\bibfnamefont{A.}~\bibnamefont{Acin}},
\bibinfo{author}{\bibfnamefont{W.}~\bibnamefont{Tittel}},
\bibinfo{author}{\bibfnamefont{H.}~\bibnamefont{Zbinden}}, \bibnamefont{and}
\bibinfo{author}{\bibfnamefont{N.}~\bibnamefont{Gisin}},
\bibinfo{journal}{J. Mod. Opt.} \textbf{\bibinfo{volume}{51}},
\bibinfo{pages}{1637} (\bibinfo{year}{2004}).

\bibitem{marcikic:02}
\bibinfo{author}{\bibfnamefont{I.}~\bibnamefont{Marcikic}},
\bibinfo{author}{\bibfnamefont{H.}~\bibnamefont{de Riedmatten}},
\bibinfo{author}{\bibfnamefont{W.}~\bibnamefont{Tittel}},
\bibinfo{author}{\bibfnamefont{V.}~\bibnamefont{Scarani}},
\bibinfo{author}{\bibfnamefont{H.}~\bibnamefont{Zbinden}}, \bibnamefont{and}
\bibinfo{author}{\bibfnamefont{N.}~\bibnamefont{Gisin}},
\bibinfo{journal}{Phys. Rev. A} \textbf{\bibinfo{volume}{66}},
\bibinfo{pages}{062308} (\bibinfo{year}{2002}).

\end{thebibliography}

\end{document}